\documentclass[%
 reprint,
 amsmath,amssymb,
 aps,
]{revtex4-2}
\newcommand\wordcount{
    \immediate\write18{texcount -sub=section \jobname.tex  | grep "Section" | sed -e 's/+.*//' | sed -n \thesection p > 'count.txt'}
(\input{count.txt}words)}
\usepackage{graphicx}% Include figure files
\usepackage{dcolumn}% Align table columns on decimal point
\usepackage{bm}% bold math
\begin{document}

\preprint{APS/123-QED}

\title{Transition from weak turbulence to\\ collapse turbulence regimes in MMT model}% Force line breaks with \\

\author{Ashleigh Simonis}
\author{Yulin Pan}%
 \email{\href{mailto:email@address.com}{yulinpan@umich.edu}}
\affiliation{%
Department of Naval Architecture and Marine Engineering, University of Michigan, Ann Arbor, Michigan 48109, USA}%

\date{\today}% It is always \today, today,

\begin{abstract}
It is well known that wave collapses can emerge from the focusing one-dimensional (1-D) Majda-McLaughlin-Tabak (MMT) model as a result of modulational instability. However, how these wave collapses affect the spectral properties and statistics of the wave field has not been adequately studied. We undertake this task by simulating the forced-dissipated 1-D MMT model over a range of forcing amplitudes. Our results show that when the forcing is weak, the spectrum agrees well with the prediction by wave turbulence theory with few collapses in the field. As the forcing strength increases, we see an increase in the occurrence of collapses, together with a transition from a power-law spectrum to an exponentially decaying spectrum. Through a spectral decomposition, we find that the exponential spectrum is dominated by the wave collapse component in the non-integrable MMT model, which is in analogy to a soliton gas in integrable turbulence.

\end{abstract}

\maketitle

%\tableofcontents

\section{\label{sec:level1}Introduction}
Wave turbulence occurs in physical systems consisting of large ensembles of weakly interacting nonlinear dispersive waves. Wave turbulence theory (WTT) provides a statistical description of the behavior of these wave systems and has rich applications in many physical contexts such as plasma physics \cite[e.g.,][]{GALTIER2000}, physical oceanography \cite[e.g.,][]{zak1967}, acoustics \cite[e.g.,][]{lvov1997} and optics \cite[e.g.,][]{PICOZZI20141}. The centerpiece of WTT is the so-called wave kinetic equation (WKE), which describes the evolution of the wave spectrum due to wave-wave interactions, and yields the Kolmogorov-Zakharov (KZ) spectra as stationary solutions \cite{Zakharov1965}. Over the decades, many efforts \cite[e.g.,][]{KOROTKEVICH2008361,pan_yue2014,Hrabski_Pan_2022,Zhang_Pan_2022,falcon2022,banks2022,zhu2022,zhu23,Simonis_Hrabski_Pan_2024} have been made to verify the WKE and KZ solutions in both numerical and experimental settings.

One model that holds a special position in the development of WTT verification is the Majda-McLaughlin-Tabak (MMT) model, which was introduced in 1997 in \cite{Majda1997} as a testbed for WTT. In \cite{Majda1997} it was found that the numerical simulation of the MMT equation yields a stationary spectrum that is significantly steeper than the KZ solution. Among several efforts to explain the discrepancy, Zakharov \cite{zakharov2001} argued that it may result from the existence of coherent structures in the wave field generated by the MMT model. In particular, it is shown in \cite{zakharov2001,ZAKHAROV20041,rumpfnewell2009} that the defocusing MMT model allows the solution of quasi-solitons, while the focusing MMT model allows wave collapses, i.e., finite-time high-amplitude singularities. In this paper, we adopt the terminology of focusing/defocusing nonlinearity \cite{cai1999} to refer to cases where the nonlinear and dispersive terms have the opposite/same sign, despite the possible confusion in the context of the MMT model as pointed out in \cite{Rumpf2013}. However, Zakharov's argument is not widely accepted, and the result of the MMT study \cite{Majda1997} has remained a mystery to the wave turbulence community for many years. This was until the recent study \cite{du_buhler_2023} (also see \cite{hrabski_pan2024verification}) for the defocusing MMT model, which clarifies that the width of inertial range, a factor ignored in previous studies, plays a critical role in the power-law spectral slope. Although the width of the inertial range realized in \cite{Majda1997} is evidently too narrow, further widening of the inertial range allows the spectrum to approach the KZ solution, irrespective of the nonlinearity level and possible coherent structures. However, it remains unclear how the spectrum behaves in the focusing MMT model and whether wave collapses affect the spectral properties.

Generally speaking, wave collapses can be induced by modulational instability, resulting in the formation of a point singularity in finite time. At the time of a collapse, both the quadratic and quartic components of the Hamiltonian surge with the total Hamiltonian conserved, which makes collapse prohibited for defocusing nonlinearity. While wave collapses and the mechanism for their formation are well studied in the nonlinear Schr\"{o}dinger equations (NLS) \cite[e.g.,][]{sulem2007nonlinear}, we highlight that the MMT model is susceptible to a different kind of instability. In \cite{Rumpf2013} it was shown that the focusing MMT equation admits a modulational instability by short-wave modulations, i.e., the wavelength of the modulation is much smaller than that of the carrier wave, in contrast to the typical Benjamin-Feir instability. Wave collapses generated from a random wave field have been investigated in several studies \cite[e.g.,][]{cai1999,cai2001,zakharov2001,ZAKHAROV20041,Rumpf2005,Rumpf2013,rumpfsheff2015}, focusing on their effect on intermittency, energy transfer mechanisms, and inception of modulational instability. However, the nonlinearity level achieved in these studies is only moderately high, with random waves still the dominating feature and the spectrum maintaining a power-law form (which is not the case with a further increase of nonlinearity, as we will show). The regime of wave collapses dominating the wave field at stronger nonlinearity has not been well understood.  

In this work, we numerically study the focusing MMT dynamics in forced-dissipated simulations covering a broad range of nonlinearity levels, from a weak wave turbulence regime to a regime where wave collapses become dominant. At a low nonlinearity level, we find a wave turbulence dominated regime with few collapses, with the spectrum consistent with the KZ solution. With the increase of nonlinearity level, we see more collapses in the field, with a flattened spectrum and departure from Gaussian statistics. At a sufficiently high nonlinearity level, the collapses become more dominant, and the spectrum transitions from power-law to exponential, together with the statistics returning to quasi-Gaussian. Through a spectral decomposition, we show that the exponential spectrum is due to the dominant collapse components, indicating a transition to a new ``collapse turbulence'' regime. This can be understood as an analogy to soliton turbulence in integrable systems, e.g., 1-D Korteweg-de Vries (KdV) and NLS equations, where exponential spectrum is also observed \cite[e.g.][]{PELINOVSKY2006425,gelash2018}. We finally show that in the collapse turbulence regime, the random wave components evolve toward a thermo-equilibrium state with reduced flux. 

\begin{figure*}[!htbp]
\includegraphics{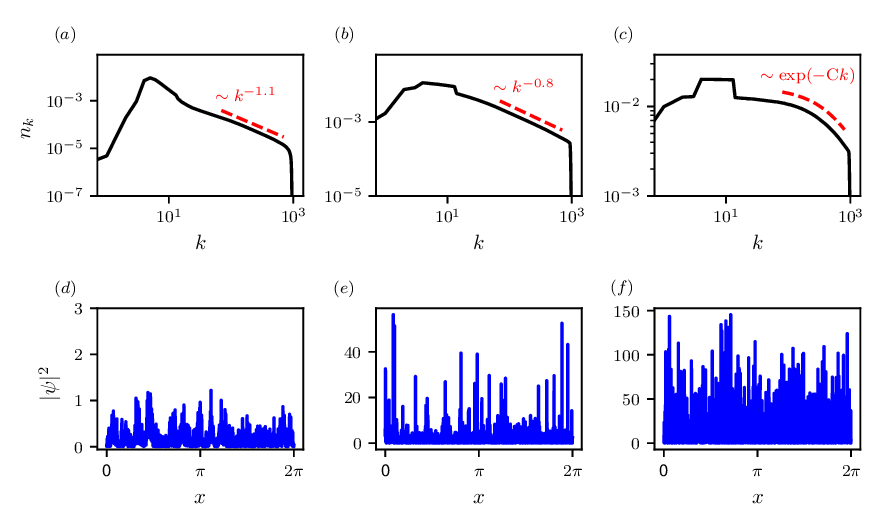}% Here is how to import EPS art
\caption{\label{fig:fig1}(top) wave action spectra and (bottom) corresponding snapshots of the $|\psi|^2$ for three levels of nonlinearity with $(a,d)$ $\epsilon = 0.03$, $(b,e)$ $\epsilon = 0.21$, $(c,f)$ $\epsilon = 0.89$.}
\end{figure*}

\section{\label{sec:level2}Numerical Procedure}

We consider the one-dimensional (1-D) MMT equation with focusing nonlinearity,

\begin{equation}
  \textnormal{i}\frac{\partial\psi}{\partial t}={\lvert{\partial _x}\rvert}^\alpha \psi-{\lvert{\partial _x}\rvert}^{\beta/4}({\big\lvert{\lvert{\partial _x}\rvert}^{\beta/4}\psi\big\rvert}^2{\lvert{\partial _x}\rvert}^{\beta/4}\psi),
  \label{MMT}
\end{equation}
where $\psi(x,t)$ is a field taking complex values and the operator $|\partial_x|^{\alpha}$ denotes the multiplication by $|k|^{\alpha}$ on each component in the spectral domain. The parameter $\beta$ controls the nonlinearity formulation and $\alpha$ controls the dispersion relation $\omega(k)=|k|^\alpha$ with $\omega$ the frequency and $k$ the wavenumber. We fix $\alpha=1/2$ and $\beta=0$, which is the same as in previous studies \cite[e.g.,][]{zakharov2001,ZAKHAROV20041,Rumpf2005,Rumpf2013,rumpfsheff2015}. The MMT equation \eqref{MMT} conserves total action $N=\int\lvert{\psi}\rvert^2dx$ and the Hamiltonian $H=H_2+H_4$ with the linear and nonlinear parts,
\begin{equation}
  \begin{aligned}
  H_{2} =\int\big\lvert{\lvert{\partial _x}\rvert}^{\alpha/2}\psi \big\rvert^2 dx,
 \\
    H_{4} =-\frac{1}{2}\int{\big\lvert{\lvert{\partial _x}\rvert}^{\beta/4}\psi\big\rvert}^4 dx.
\end{aligned}
\end{equation}

Each numerical simulation is performed with 4096 modes, which corresponds to a maximum wavenumber of 1024 after dealiasing, on a periodic domain of $L = 2\pi$. We start simulations of \eqref{MMT} from a low-amplitude background of random waves as initial conditions, and let the field evolve into a stationary state under forcing and dissipation. The forcing is in white-noise form, given by
\begin{equation}
  F =
    \begin{cases}
      F_{r}+\textnormal{i}F_{i}, & 4\le k \le 13,\\
      0, & \text{otherwise},
    \end{cases}
    \label{F}
\end{equation}
with $F_r$ and $F_i$ independently drawn from a Gaussian distribution $\mathcal{N}(0, \sigma^2)$. We use a broad range of $\sigma\in [0.037, 3.41]$ in simulations to ensure that the nonlinearity level achieved covers the range of interest. The dissipation is imposed with the addition of two hyperviscosity terms
\begin{equation}
  \left.\begin{aligned}
  D_{1} =
    \begin{cases}
      -\textnormal{i}\nu_{1}\hat{\psi}_{k}, & k\ge 900,\\
      0, & \text{otherwise},
    \end{cases} \\
    D_{2} =
    \begin{cases}
      -\textnormal{i}\nu_{2}\hat{\psi}_{k}, & k\le 4,\\
      0, & \text{otherwise},
    \end{cases} 
    \label{D}
\end{aligned}\right\}
\end{equation}
at small and large scales, respectively. The MMT model admits an inverse cascade; therefore, the addition of large-scale dissipation is necessary to prohibit the accumulation of energy at these scales. The dissipation coefficients are set to $\nu_{1}=10^{-14}(k-900)^8$ and $\nu_{2}=3k^{-4}$ for all numerical experiments. %Additional details on the numerical schemes used for the simulations can be found in \cite{hrabski_pan2020,Hrabski_Pan_2022,Simonis_Hrabski_Pan_2024}.

\section{\label{sec:level3}Results}

\subsection{\label{sec:spec_form}Spectral properties \& statistics}
We define $\epsilon=H_4/H_2$ in the stationary state as a measure of the nonlinearity level of the wave field and the wave action spectrum $n_k=\langle\hat{\psi}_k\hat{\psi}^*_k\rangle$ with $\hat{\psi}_k$ the Fourier transform of $\psi$ and the angle brackets denoting an ensemble average. Figure 1 shows the wave action spectrum $n_k$, as well as the corresponding wave field for three very different nonlinearity levels ranging from $\epsilon\in[0.03,0.89]$. At a low nonlinearity level (Fig. \ref{fig:fig1}(a), (d)), there are few collapses in the field and the spectrum exhibits a power-law form with a slope close to the KZ prediction of $\gamma=-1$. With an increase of nonlinearity (Fig. \ref{fig:fig1}(b), (e)), we see more collapses emerging from the field and the power-law spectrum becomes flatter than the KZ prediction. These observations are consistent with previous studies \cite{cai1999,cai2001}. At very high nonlinearity exceeding that of previous studies (Fig. \ref{fig:fig1}(c), (f)), the field becomes saturated with collapses, and the wave action spectrum departs from a power-law and tends toward an exponential form. We remark that such an exponential spectrum is similar to that of a soliton gas in integrable turbulence \cite[e.g.,][]{PELINOVSKY2006425,gelash2018,gelash2019,Redor2019}, where a large number of solitons exist with a background of random waves \cite{El_2021}. For non-integrable systems, the only known example (to the authors) of the exponential spectrum is realized in discrete NLS \cite{Rumpf2004}, where a state of coexistence of waves and localized excitations is involved and termed “two-species gas”. For continuous non-integrable systems, the authors are not aware of other examples, and the result for the MMT model is therefore a new discovery.

\begin{figure}
\includegraphics{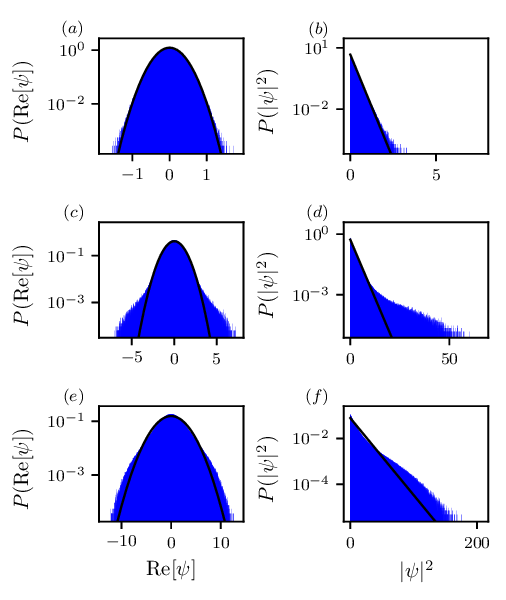}
\caption{Probability distributions functions of (left) Re[$\psi$] and (right) $|\psi|^2$ for for three levels of nonlinearity with $(a,b)$ $\epsilon = 0.03$, $(c,d)$ $\epsilon = 0.21$, $(e,f)$ $\epsilon = 0.89$. Each panel is fitted with a Gaussian or exponential distribution of the same mean and standard deviation (solid line).}
\label{fig:fig2}
\end{figure}

We further investigate the wave statistics at different nonlinearity levels. Figure \ref{fig:fig2} shows the probability distribution functions (PDFs) of $\textnormal{Re}[\psi]$ and squared amplitude $|\psi|^2$ at the same nonlinearity levels as those in Fig. \ref{fig:fig1}. At a low nonlinearity level (Fig. \ref{fig:fig2}(a), (b)), we see that the histogram of $\textnormal{Re}[\psi]$ is well fitted by a Gaussian distribution and $|\psi|^2$  follows an exponential distribution. At a higher level of nonlinearity (Fig. \ref{fig:fig2} (c), (d)), deviations from the Gaussian and exponential distributions are observed with fatter tails, indicating increased intermittency of the system. These behaviors are consistent with those of previous studies for the MMT model \cite{Rumpf2005,rumpfsheff2015,COUSINS201448}, as well as in the general observation of intermittency in wave turbulence \cite[e.g.,][]{lvov_naz04,falcon2007,Falcon_2010,ALBERELLO201981}. With a further increase of nonlinearity to the level where wave collapses become dominant (Fig. \ref{fig:fig2}(e), (f)), we interestingly find that the statistics of $\textnormal{Re}[\psi]$ return to be close to Gaussian (and $|\psi|^2$ to exponential). This is likely because when collapses are dominant, they become the main, rather than intermittent, feature of the field. Therefore, the intermittency and non-Gaussian tail of the PDF have to be reduced.

We note that an extended self-similarity analysis (ESS) can also be performed following the procedure in \cite{chibbaro2017}. Assuming structure functions $S_p\sim r^{\xi_p}$ and $S_2\sim r^{\xi_2}$, the purpose of this analysis is to compare $\xi_p/\xi_2$ from numerical data with scale-invariant scaling $p/2$. Our results of this analysis (not shown) indicate that as $\epsilon$ increases from 0.03 to 0.21, $\xi_p/\xi_2$ departs from $p/2$ due to intermittency, consistent with \cite{chibbaro2017}. As $\epsilon$ increases further to 0.89, we observe that $\xi_p/\xi_2$ returns to the $p/2$ scale. However, it must be emphasized that at high nonlinearity, the spectrum is not power-law, violating the scaling of the structure functions as the basis of ESS. Therefore, this result needs to be interpreted with caution, which we choose not to stress in this paper. 

\begin{figure}
\includegraphics{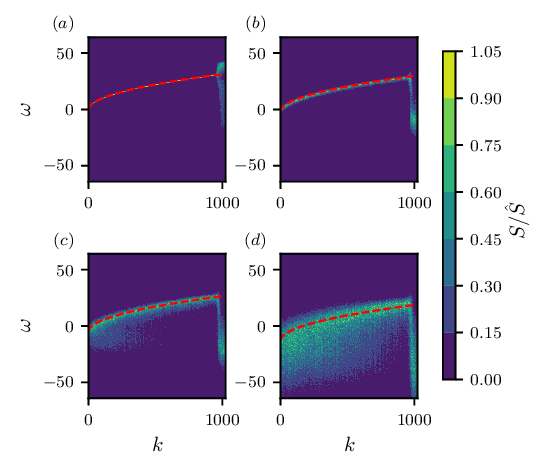}
\caption{The wave number-frequency spectra $S/\hat{S}$ for nonlinearity level $(a)$ $ \epsilon = 0.03$, $(b)$ $\epsilon = 0.21$, $(c)$ $\epsilon = 0.48$, and $(d)$ $\epsilon = 0.89$. The red dashed lines indicate the renormalized dispersion relation $\tilde{\omega}(k)$. }
\label{fig:fig3}
\end{figure}

\begin{figure}
\includegraphics{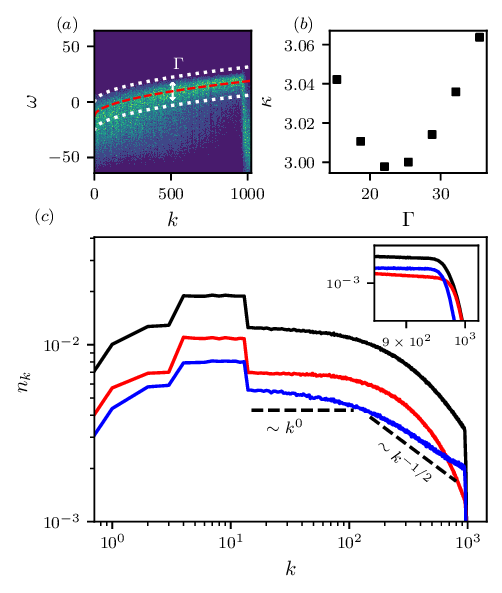}
\caption{Spectral decomposition for the case with $\epsilon=0.89$. (a) an illustration of the broadening parameter $\Gamma$ for defining the wave component; (b) the kurtosis $\kappa$ as a function of $\Gamma$; (c) the decomposed wave action spectrum with the total spectrum (solid black), the wave component spectrum (solid blue), and the collapse component spectrum (solid red). The Rayleigh-Jeans spectra with $n_k\sim k^0$ and $n_k\sim k^{-1/2}$ are indicated by the dashed black lines. A zoomed-in view for the high-wavenumber region of the spectra is included as an inset.}
\label{fig:fig4}
\end{figure}

\subsection{\label{sec:k-omega}Wavenumber-frequency spectrum \& spectral decomposition}
We next examine the wavenumber-frequency spectrum at different nonlinearity levels, as plotted in Fig. \ref{fig:fig3}. Specifically, we have plotted the normalized spectrum $S/\hat{S}$, where $S$ is the standard wavenumber-frequency spectrum and $\hat{S} = \textnormal{max}_{\omega}S(k,w)$. In this way, the spectral behavior at each $k$ (especially large $k$) can be elucidated. Also shown in Fig. \ref{fig:fig3} are the renormalized dispersion relation curves $\tilde \omega(k)=\omega(k)-2\sum_{k_1}{|\hat{\psi}_{1}|^2}$ \cite{Nazarenko2011} as dashed lines. At low nonlinearity (Fig. \ref{fig:fig3}(a)), we see that the spectral intensity aligns well along the dispersion relation curve, suggesting the dominance of random waves in the field consistent with WTT. With the increase of the nonlinearity level (Fig. \ref{fig:fig3}(b),(c)), we see spectral broadening around the dispersion relation curve, as well as the emergence of components below the dispersion relation curve, especially in Fig. \ref{fig:fig3}(c). These are exact representations of the collapses that do not satisfy the dispersion relation. At high nonlinearity (Fig. \ref{fig:fig3}(d)), we see that the collapse component becomes more dominant, shown as signals that fill a large area in the $\omega$-$k$ space. We also note that in all figures, the random wave components follow the renormalized dispersion relation $\tilde \omega(k)$ (instead of $\omega(k)=k^{1/2}$), which is more clearly seen at higher nonlinearity. 

\begin{figure*}[!htbp]
\includegraphics{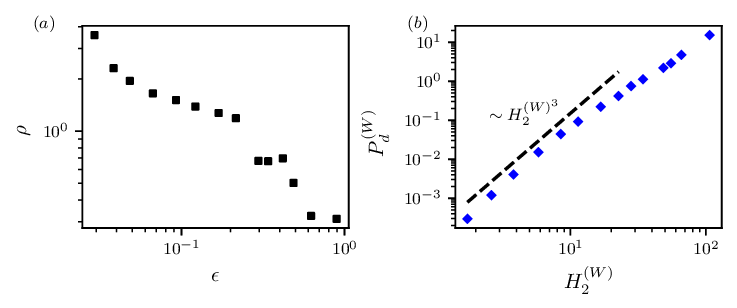}
\caption{(a) the ratio $\rho$ as a function of nonlinearity level $\epsilon$; (b) energy flux $\bar{P_d}^{(W)}$ as a function of $H_{2}^{(W)}$ for the wave component, with WTT prediction $\bar{P_d}^{(W)}\sim H^{(W)^3}_2$ denoted by the dashed line.}
\label{fig:fig5}
\end{figure*}
We further decompose the field at high nonlinearity $\epsilon=0.89$ into wave and collapse components using a variation of the method developed in \cite{LAURIE2012121}. Generally speaking, the method decomposes the wave and collapse components according to their proximity to the dispersion relation curve. Our application of the method is described in Fig. \ref{fig:fig4}. We first choose a broadening parameter $\Gamma$ (see Fig. \ref{fig:fig4}(a)) such that the spectral content within $\tilde \omega(k) \pm \Gamma/2$ satisfies a Gaussian distribution. This can be achieved by measuring kurtosis $\kappa$ as a function of $\Gamma$ as in Fig. \ref{fig:fig4}(b), and choosing the value of $\Gamma$ for which $\kappa=3$. For our case, the optimal $\Gamma$ is approximately 20-25, and we use 25 for this study. Figure \ref{fig:fig4}(a) shows that such a choice of $\Gamma$ roughly incorporates the major spectral content around the dispersion relation curve. We can then define the random wave component as the spectral content within $\tilde \omega(k) \pm \Gamma/2$, and the collapse component as the rest. Figure \ref{fig:fig4}(c) shows the decomposed wave action spectra for both the wave and the collapse components. We see that the exponential total spectrum indeed results from the collapse component that is dominant for most wavenumbers, with its spectrum taking an exponential form. On the other hand, the spectrum of the random wave component remains a power-law form close to a thermo-equilibrium state (see fitting with Rayleigh-Jeans spectrum in Fig. \ref{fig:fig4}(c)), indicating that the energy flux by random waves is suppressed in the collapse-dominant regime.

We next perform a more detailed analysis of the energy flux mechanism of the system. As mentioned in \cite{cai1999,cai2001,Rumpf2005}, there are two mechanisms of energy cascade in the focusing MMT model: one local transport in $k$-space from wave-wave interactions of random waves, and the other nonlocal transport in $k$-space from the formation of small-scale wave collapses. Therefore, the total energy flux results from the summation of the two mechanisms. Our goal is to understand the relative importance of the two mechanisms at different levels of nonlinearity. Considering that the energy flux from the two mechanisms is equal to the dissipation of random waves and collapses, respectively, we define a ratio
\begin{equation}
  \rho = \frac{\bar{P_d}^{(W)}}{\bar{P_d}^{(C)}}
  \label{rat}
\end{equation}
where $\bar{P_d}^{(W)}$ and $\bar{P_d}^{(C)}$ are the dissipations of wave and collapse components, calculated as
\begin{equation}
  \bar{P_d}^* = \sum_{k>k_d=900}-2\nu_1\omega_k n_k^*,\; *=(W),(C),
  \label{pd}
\end{equation}
with $n_k^*$ decomposed as in Fig. \ref{fig:fig4}(c). 

Figure \ref{fig:fig5}(a) plots the ratio $\rho$ as a function of nonlinearity level $\epsilon$. We see a significant reduction in $\rho$ with the increase of $\epsilon$, indicating that the fraction of energy flux from random waves decreases substantially with the increase of nonlinearity. According to this result, we conclude that the system behaves in the following way: As nonlinearity increases, the energy flux due to wave-wave interactions grows following the WTT prediction of $\bar{P_d}^{(W)}\sim H^{(W)^3}_2$ up to moderately high nonlinearity. Beyond this point, the energy flux from the wave component becomes lower than the WTT prediction, as seen in Fig. \ref{fig:fig5}(b). This is consistent with the flattened spectrum toward the thermal-equilibrium state. This transition point occurs at approximately $\rho \lesssim O(1)$, indicating a significant portion of the contribution to the energy flux of the collapse component. Meanwhile, the energy flux from the collapse component increases much faster than that from waves, leading to a decreased value of $\rho$ with an increase in nonlinearity as seen in Fig. \ref{fig:fig5}(a).

\section{\label{sec:level3}Conclusion}
In this work, we numerically study the spectral properties and statistics from the forced-dissipated 1-D focusing MMT equation, which admits wave collapses due to modulational instability. Our work covers a broader range of forcing (thus nonlinearity levels) than previous works, and therefore reveals the physics at a sufficiently high nonlinearity level when the wave collapses become the dominant feature. We show that as nonlinearity increases toward this collapse-dominant regime, the spectrum departs from a power-law form and tends toward an exponential form. In the meantime, the system surpasses the intermittent regime characterized by strongly non-Gaussian statistics and recovers the quasi-Gaussian statistics. The exponential spectrum resembles what is typically seen for soliton gas in integrable turbulence, but is now realized in a non-integrable system. Through a spectral decomposition method,  we show that the nature of the exponential spectrum can indeed be attributed to the dominant collapse component. With the presence of these coherent structures, the energy flux from wave-wave interactions is reduced from the prediction by WTT with the spectrum of wave component tending toward a thermo-equilibrium state. 

\begin{acknowledgments}
The authors thank Dr. Alexander Hrabski for the insightful discussion during this work. This research was supported by the Simons Foundation (Award ID \#651459). The computation was performed on the Great Lakes HPC Cluster provided by Advanced Research Computing (ARC) at the University of Michigan, Ann Arbor. 
\end{acknowledgments}

\nocite{*}

\bibliography{mmt_collapse_simonis-pan}% Produces the bibliography via BibTeX.

\end{document}